\documentclass[twocolumn]{aastex631}

\usepackage{makecell}

\usepackage{graphicx}
\usepackage{natbib}
\usepackage{amsmath}
\usepackage{mathtools}
\usepackage{subcaption}
\usepackage{url}

\usepackage{newtxtext,newtxmath}

\newcommand{\Msun}{M$_{\odot}$}

\shorttitle{Cygnus X-1 X-ray polarimetry with {\em Astrosat}}
\shortauthors{Chattopadhyay et al.}

\begin{document}

\title{High hard X-ray  polarization in Cygnus X-1 confined to the intermediate hard state:  evidence for a variable jet component}

\correspondingauthor{Tanmoy Chattopadhyay}
\email{tanmoyc@stanford.edu}

\author{Tanmoy Chattopadhyay}
\affiliation{Kavli Institute of Particle Astrophysics and Cosmology, Stanford University \\
452 Lomita Mall, Stanford, CA 94305, USA}

\author{Abhay Kumar}
\affiliation{Physical Research Laboratory \\
Thaltej, Ahmedabad, Gujarat 380009, India}

\author{A. R. Rao}
\affiliation{Inter-University Center for Astronomy and Astrophysics\\
Pune, Maharashtra-411007, India}

\author{Yash Bhargava}
\affiliation{Department of Astronomy and Astrophysics, Tata Institute of Fundamental Research, \\1 Homi Bhabha Road, Colaba, 400005 Mumbai, India}
\affiliation{Inter-University Center for Astronomy and Astrophysics\\
Pune, Maharashtra-411007, India}

\author{Santosh V. Vadawale}
\affiliation{Physical Research Laboratory \\
Thaltej, Ahmedabad, Gujarat 380009, India}

\author{Ajay Ratheesh}
\affiliation{INAF - IAPS, Via Fosso del Cavaliere 100, I-00133 Rome, Italy}

\author{Gulab Dewangan}
\affiliation{Inter-University Center for Astronomy and Astrophysics\\
Pune, Maharashtra-411007, India}

\author{Dipankar Bhattacharya}
\affiliation{Inter-University Center for Astronomy and Astrophysics\\
Pune, Maharashtra-411007, India}
\affiliation{Department of Physics, Ashoka University \\ Rai, Sonipat, Haryana-131029, India}

\author{N. P. S. Mithun}
\affiliation{Physical Research Laboratory \\
Thaltej, Ahmedabad, Gujarat 380009, India}

\author{Varun Bhalerao}
\affiliation{Indian Institute of Technology Bombay\\
Powai, Mumbai, Maharashtra 400076, India}

\begin{abstract}
   
Cygnus X-1, the well-known accreting black hole system,  
exhibits several observational features hinting at an intricate interplay between the accretion disk, its atmosphere known as the corona and the putative relativistic jet. 
It has been extensively studied using all available observational methods, including using the newly available technique of sensitive X-ray polarimetry.
X-ray polarization characteristics are distinct for coronal and jet emissions. 
The low  X-ray polarization measured below $\sim$100 keV is understood as arising from the corona. In contrast, the high polarization measurements reported above $\sim$400 keV required a separate jet-dominated spectral component, which spectroscopy does not demonstrate conclusively. 
Here we report precise polarization measurements in the 100--380 keV region made during three different sub-classes of spectral states of the source using the CZTI instrument onboard {\em AstroSat}. A high polarization (23$\pm$4 \%) is found mainly in the Intermediate Hard State of the source, and the energy-resolved measurements smoothly connect the coronal and the jet regimes.
When high polarization is observed, the simultaneous spectral data hints at a separate power law component above 100 keV. We examine the possible sources of this energy-dependent high polarization in Cygnus~X-1. 

\end{abstract}



\section{Introduction} \label{sec:intro}

Cygnus~X-1, a high-mass X-ray binary (HMXB) system, is one of the earliest known X-ray sources, harboring a 21.2$\pm$2.2~\Msun\ black hole in a 5.6-day orbit with a 40.6$^{+7.7}_{-7.1}$~\Msun\ star, and located
at a distance of 2.22$^{+0.18}_{-0.17}$~kpc from us \citep{miller21}.
Unlike most other X-ray sources, Cygnus~X-1 is persistent and has been extensively studied across 
almost the entire electromagnetic spectrum over the last five decades.
The source displays state 
transitions between the thermal disk dominated soft state and the
hard state with a power-law dominated spectrum \citep{sunyaev79_cgx1,ebisawa96,gierlinski97_cgx1,cui97_cgx1,salvo01_cgx1,mcconnel02_cgx1}. It is also detected 
in radio wavelengths during different spectral states, thought to originate in relativistic jets \citep{stirling01_cgx1,gallo03_cgx1,fender06_cgx1,Wilms07_cgx1}. The source is also detected in GeV emission, also thought to be associated with the relativistic jets \citep{zanin16}.
Cygnus~X-1 is one of the brightest X-ray sources in the hard state, and the hard X-ray emission is attributed mainly to Compton scattering from a hot corona. 
\citet{mcconnel02_cgx1} reported detection of high-energy emission above Comptonization cut-off which was explained in terms of hybrid Comptonization.
More recently, this additional component in the hard state spectrum \citep{bel06_cgx1,Jourdain14_cgx1} has been interpreted as power-law emission from an optically thin jet based on simultaneous multi-wavelength studies \citep{rahoui11_cgx1}. 
Detailed modeling of the broadband spectral energy distribution (SED) of Cygnus~X-1 in the hard state requires consideration of jet emission to account for the soft-gamma ray observations \citep{zdziarski14}. It has been suggested that under certain conditions, the jet emission may contribute significantly in hard X-rays as well \citep{malyshev13_cgx1,russell14_cgx1,Kantzas20}, similar to a few other black hole sources \citep{vadawale01,markoff01,vadawale03}.
However, the extent to which the jet emission can contribute to hard X-rays continues to be debated \citep{zdziarski14}.
Hard X-ray polarization measurements offer a unique possibility to distinguish
between emissions arising in the corona and the jet. However, hard X-ray
polarization measurements are challenging to carry out, and so far only weak hints are available of polarization in hard X-rays \citep{chattopadhyay21_review}.

The first attempt to explore the polarization properties of Cygnus~X-1 dates way back to 1970s when a Bragg polarimeter onboard The Eighth Orbiting Solar Observatory ({\em OSO 8}) placed an upper limit of a few percent at 2.6 keV \citep{long80}. Subsequently, there have been attempts to measure polarization of the source in hard X-rays, both in the coronal regime (a few tens of keV to $\sim$100 keV) and the suspected jet regime (above 100 keV) \citep[see][for a summary]{chattopadhyay21_review}. 
Recently, \citet{krawczynski22_ixpe} reported a precise measurement of polarization of Cygnus~X-1 in the hard state using Imaging X-ray Polarimetry Explorer ({\em IXPE}) in 2--10 keV band. 
They found a polarization degree (PD) of 4.0$\pm$0.2 \% with an increasing trend in polarization with energy. The polarization angle (PA) is -20.7$\pm$1.4$^\circ$ (from the local north towards northeast in clockwise direction) and aligns with the outflowing radio jet. In this paper, we convert all the reported PAs in 0--180$^\circ$ range from the local north in anti-clockwise direction. According to this convention, the measured IXPE PA is 159.3$^\circ$. These results suggest that the X-ray coronal plasma is extended in the plane of the accretion disk. 
IBIS and SPI instruments onboard the INTErnational Gamma-Ray Astrophysics Laboratory ({\em INTEGRAL}) independently measured high polarization for this source at $\sim$65 \% with a polarization angle of 44$^\circ$ at energies above 400 keV \citep{laurent11,jourdain12}. These results were interpreted as the jet origin of the photons, further corroborated by the spectroscopic analysis showing two distinct spectral components, a thermal Comptonization component at energies below 200 keV and a power law component beyond 200 keV, supposedly, due to synchrotron radiation from the jet.
However, \citet{zdziarski14} modeled wide band spectral energy distribution
of Cygnus~X-1 spanning from radio to MeV and suggested that for a realistic set of model parameters, contribution of the jet emission in X-rays is likely negligible.  
Later observations by the Polarized Gamma-ray Observer ({\em PoGO+}), a dedicated balloon-borne hard X-ray polarimeter sensitive in 19--181 keV, found the source to be unpolarized in the hard state. 
They placed an upper limit of 5.6 \% \citep{chauvin18a}.
They also estimated upper limits for polarization from the jet component of around 5--10 \% \citep{chauvin18b}.

These findings enhance the tension between the low energy polarization measurements and the high polarization found above 250 keV by {\em INTEGRAL}, requiring a synchrotron emission component. A detailed polarimetric study of the source in 100--500 keV region (the energy range in which the coronal and the jet components could have similar contribution) can confirm any separate jet component in the hard X-rays. Since the radio emission, believed to be originating from the jet, is known to change between different spectral states of Cygnus~X-1, 
it is essential to have hard X-ray polarization measurements in different spectral states to decipher the underlying emission mechanisms. However, such state-dependent hard X-ray polarization measurements have not been possible so far.

Cadmium Zinc Telluride Imager (CZTI) is a moderately sensitive hard X-ray polarimeter in 100--380 keV energy range. 
The polarization information is obtained by accurately identifying the Compton scattered events in the CZTI plane, which modulate the azimuthal angle distribution if the incident radiation is polarized.
The capability of CZTI as a polarimeter has been demonstrated both in the laboratory before the launch of {\em AstroSat} \citep{vadawale15,chattopadhyay14} 
and in space with the measurement of polarization of Crab \citep{vadawale17}. Polarization measurements for a large sample of Gamma-ray Bursts (GRBs) have also been reported by \citet{chattopadhyay19} and \citet{chattopadhyay22_grb}. 
CZTI polarimetry range (100--380 keV) bridges the gap between the {\em PoGO+} estimate and {\em INTEGRAL} measurements and, therefore, can contribute significantly to understanding the emission mechanism in this energy range.
If there is indeed a transition in the emission mechanism from corona to jet, that can be effectively probed by studying the energy-resolved polarization properties of the source with CZTI. 
For this study, we made three long targeted observations of Cygnus~X-1 after the source transitioned to hard state (hereafter ID2992, ID4646, and ID5146). We did a detailed polarization analysis of these three observations using the Compton events in CZTI. 
  
In Sec. \ref{sec:obs}, we give the details of the observations along with their spectral state determination. The polarization results are discussed in Sec. \ref{sec:pol_results} followed by a brief description of the spectroscopic analysis and results in Sec. \ref{sec:spec_results}. In Sec. \ref{sec:discussion}, we discuss the results in the context of coronal and jet contribution to the global emission of Cygnus~X-1 in different spectral states. 


\section{AstroSat Observation of Cygnus~X-1}\label{sec:obs}

Since the launch of {\em AstroSat}, Cygnus~X-1 has been observed on several occasions. Many of them, however, are of short exposures, and some are in the soft state with very low hard X-ray flux, not suitable for polarization analysis. 
Hence, during the last few observations cycles, three long ($>$200 ks) observations ({\em AstroSat} observations ID2992, ID4646, and ID5146), triggered by the transition of the source from soft to hard state, were undertaken. 
Details of the source and blank sky observations used for polarization analysis are given in Table \ref{tab:obs_res}.

\begin{tiny}
\begin{table*}
\begin{center}
\begin{footnotesize}
\caption{ Summary of the Cygnus~X-1 and the blank sky observations}\label{tab:obs_res}

\begin{tabular}{llllllllllll}
\hline
Observation$^*$   & Exposure  & Date of & Power law   & Flux $\times$10$^{-10}$  [3]  & Energy &PD & PA [4]  & Source & Background & Modulation & MDP \\
ID&(ks)&observation&index [2]&erg s$^-1$ cm$^-2$ & range & (\%)&&Counts& Counts&Factor & (\%)\\
&&&(30--100 keV)& (22--100 keV) & (keV)& &&&&\\

\hline
9000002992 & 257   & 15--21/06/2019  & 2.06$\pm$0.01&90.96$\pm$0.22&100--380& 23$\pm$4  & 56$\pm$11$^\circ$  & 277662 & 3440140 & 0.075$\pm$0.015 & 7\\
&&&&&100--175&$<$15  & & 69565 & 882723 & 0.038$\pm$0.030 & 13\\
&&&&&175--230&26$\pm$6  &48$\pm$12$^\circ$  & 98182 & 1227443 & 0.103$\pm$0.025 & 11\\
&&&&&230--380&39$\pm$9  & 59$\pm$11$^\circ$  & 109917 & 1329973& 0.116$\pm$0.023 & 12 \\
\hline
 9000004646 & 169  & 16--21/08/2021 & 2.14$\pm$0.02&   65.47$\pm$0.22&100--380& $<$12 &  & 137047 & 2305872 &   0.051$\pm$0.046& 12 \\
 \hline
 9000005146 & 303  & 15--23/05/2022 & 1.75$\pm$0.01&126.66$\pm$0.18&100--380& $<$7 &  & 220164 & 3872266 &  0.070$\pm$0.033 & 9  \\
 \hline
 9000002210 [1] & 207  &  03/07/2018 & --- & --- & --- \\
\hline
\end{tabular}\\

\begin{flushleft}
    $[1]$ {Blank sky observation for the measurement of polarimetric background.}\\
    $[2]$ {errors are for 1$\sigma$ level.}\\
    $[3]$ {errors are for 1$\sigma$ level.}\\
    $[4]$ {1$\sigma$ errors. The upper limits are computed at 99 \% level.}\\
    $^*$ Background and source RA/DEC: 203.65/37.91 and 299.59 /35.20, respectively.
    \end{flushleft}

\end{footnotesize}
\end{center}
\end{table*} 
\end{tiny}

\begin{figure}[h!]
	\centering
	\includegraphics[scale=0.3]{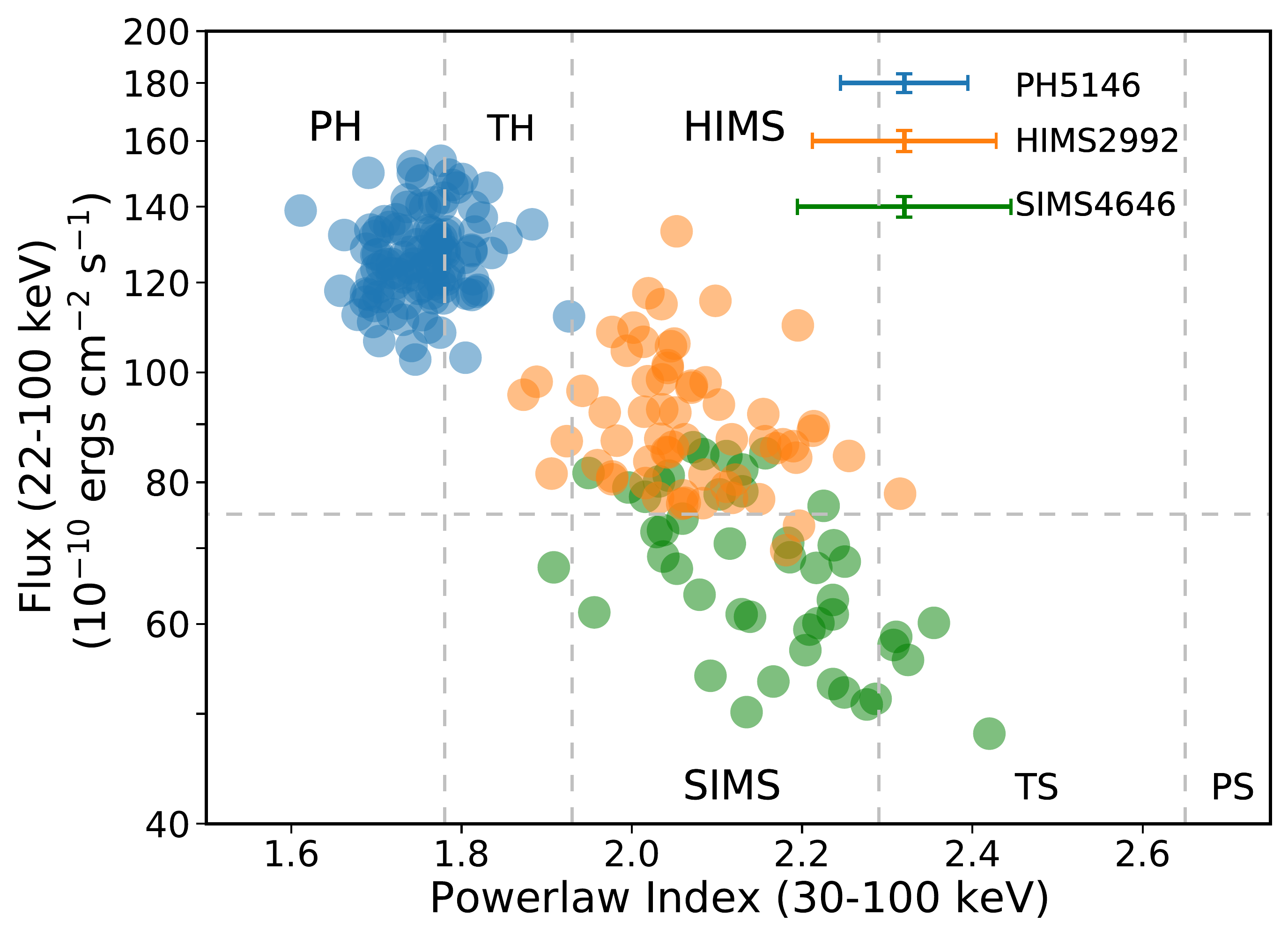}
\caption{Spectral states of Cygnus~X-1 for the three {\em AstroSat} CZTI observations. We fit the 30--100 keV CZTI mask-weighted spectrum for each orbit and measure the flux in 22--100 keV. The distribution of the fitted power law indices and the flux values are shown here. Each data point represents one orbit ($\sim$90 min long). The vertical lines are the power law index boundaries separating different spectral states, whereas the horizontal line separates the hard and the soft states as defined by \citet{Lubinski20}, thus segregating the index-flux plane into six spectral states - PH: pure hard, TH: transitional hard, HIMS: intermediate hard, SIMS: intermediate soft, TS: transitional soft, PS: pure soft. We find that ID5146 is a pure hard state (PH5146), whereas ID2992 and ID4646 belong to intermediate hard (HIMS2992) and intermediate soft (SIMS4646) states, respectively. The typical errors are shown in the legend. }
	\label{fig:states}
\end{figure}

To identify the specific subclass of the spectral state, we followed a method described by  \citet{Lubinski20}.
Spectral analysis was carried out in 30--100 keV energy range by fitting a \texttt{powerlaw} to the orbit-wise spectra, and a distribution of the orbit-wise fitted spectral index and flux (22--100 keV) was obtained for each of the three observations.
From the distributions, we classify ID2992, ID4646, and ID5146 as Intermediate Hard State (HIMS), Intermediate Soft State (SIMS), and Pure Hard state (PH), respectively as shown in Figure \ref{fig:states}. 
Hereafter, we denote these observations as HIMS2992, SIMS4646, and PH5146, respectively. Details of the method for spectral state determination can be found in supplementary material \ref{app:spec}. 

\section{Polarization analysis results} \label{sec:pol_results}
\begin{figure*}
	\centering
 \begin{subfigure}{.45\textwidth}
	\includegraphics[width=\linewidth]{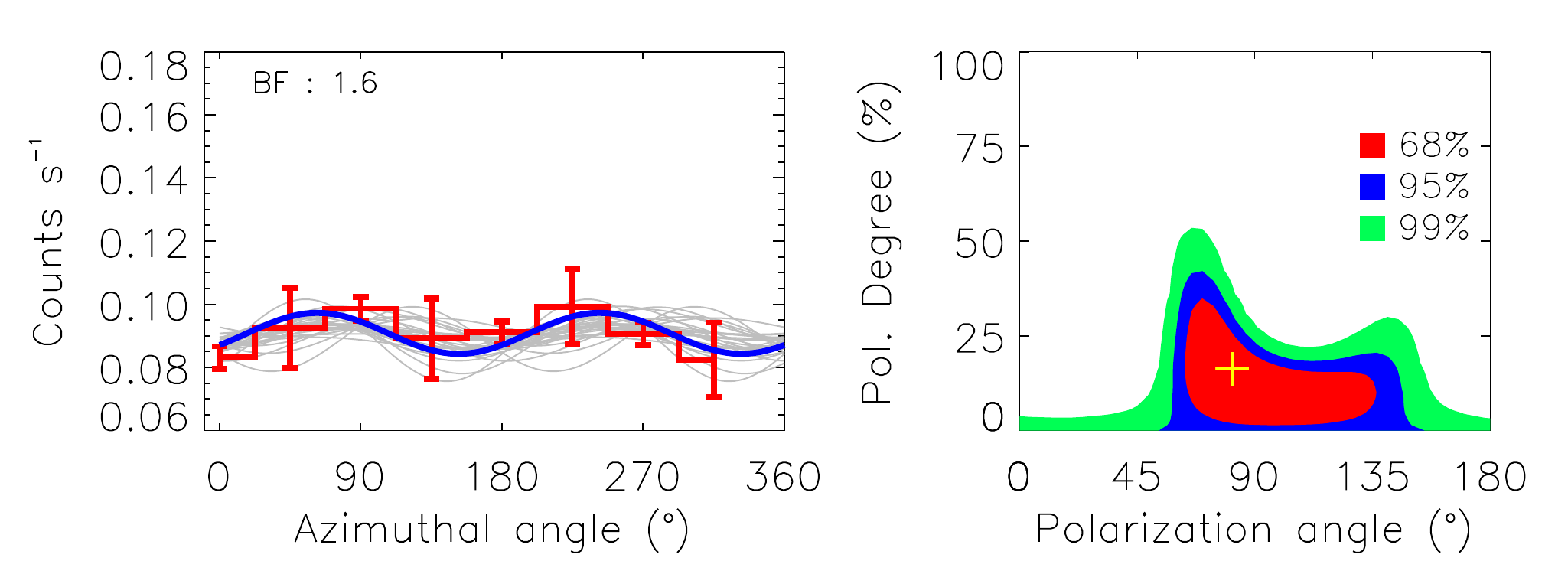}
     \caption{(a) PH5146: 100--380 keV}
    \end{subfigure}
    \begin{subfigure}{.45\textwidth}
	\includegraphics[width=\linewidth]{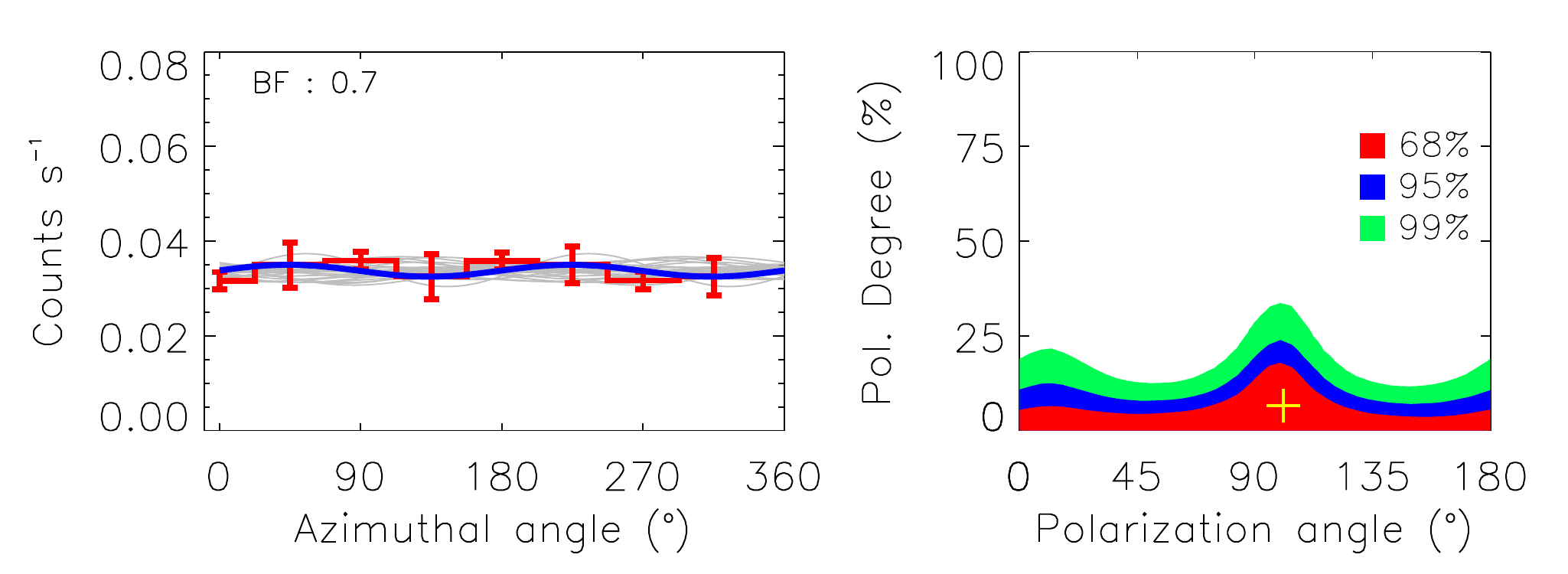}
     \caption{(d) HIMS2992: 100--175 keV}
    \end{subfigure}
     \begin{subfigure}{.45\textwidth}
		\includegraphics[width=\linewidth]{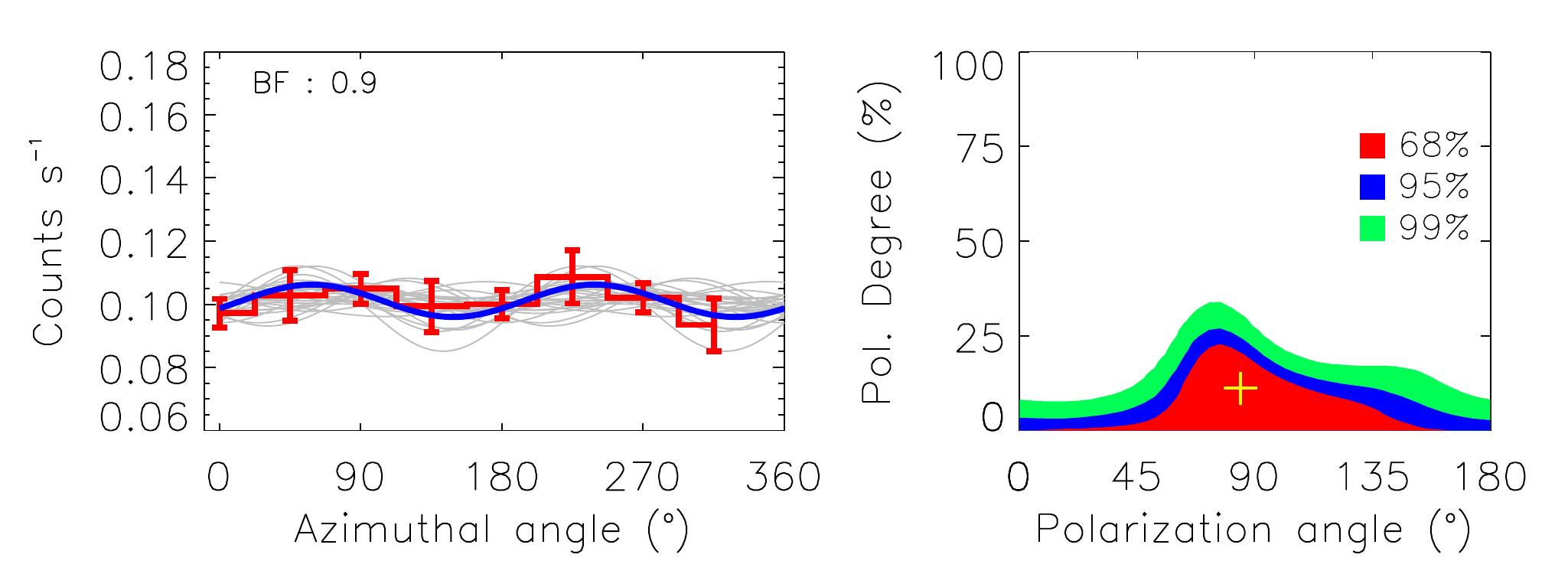}
  \caption{(b) SIMS4646: 100--380 keV}
    \end{subfigure}
    \begin{subfigure}{.45\textwidth}
	\includegraphics[width=\linewidth]{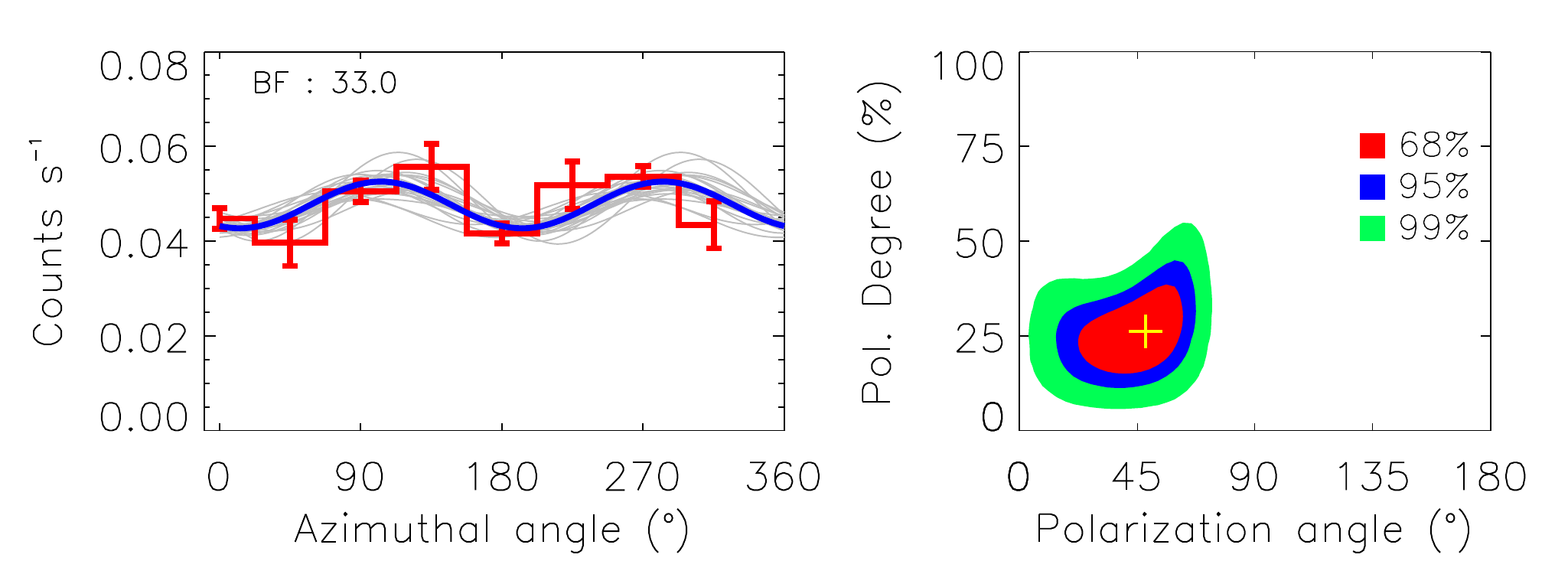}
     \caption{(e) HIMS2992: 175--230 keV}
    \end{subfigure}
     \begin{subfigure}{.45\textwidth}
		\includegraphics[width=\linewidth]{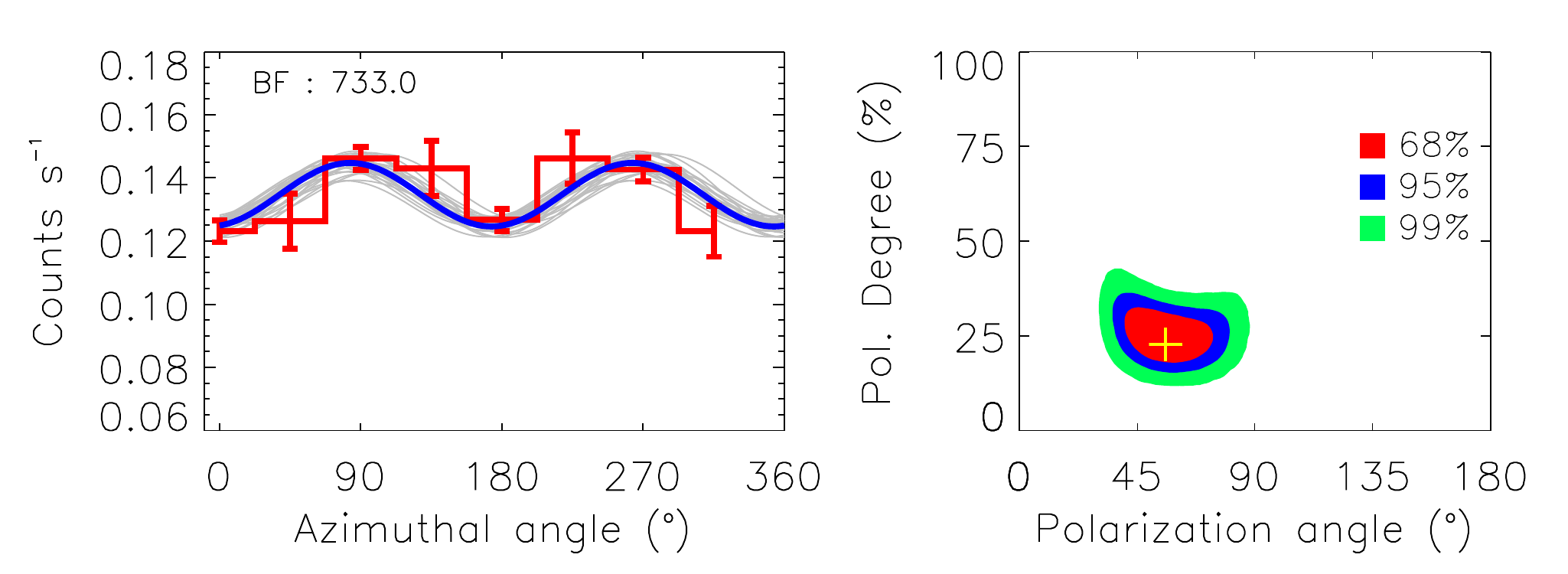}
  \caption{(c) HIMS2992: 100--380 keV}
    \end{subfigure}
    \begin{subfigure}{.45\textwidth}
	\includegraphics[width=\linewidth]{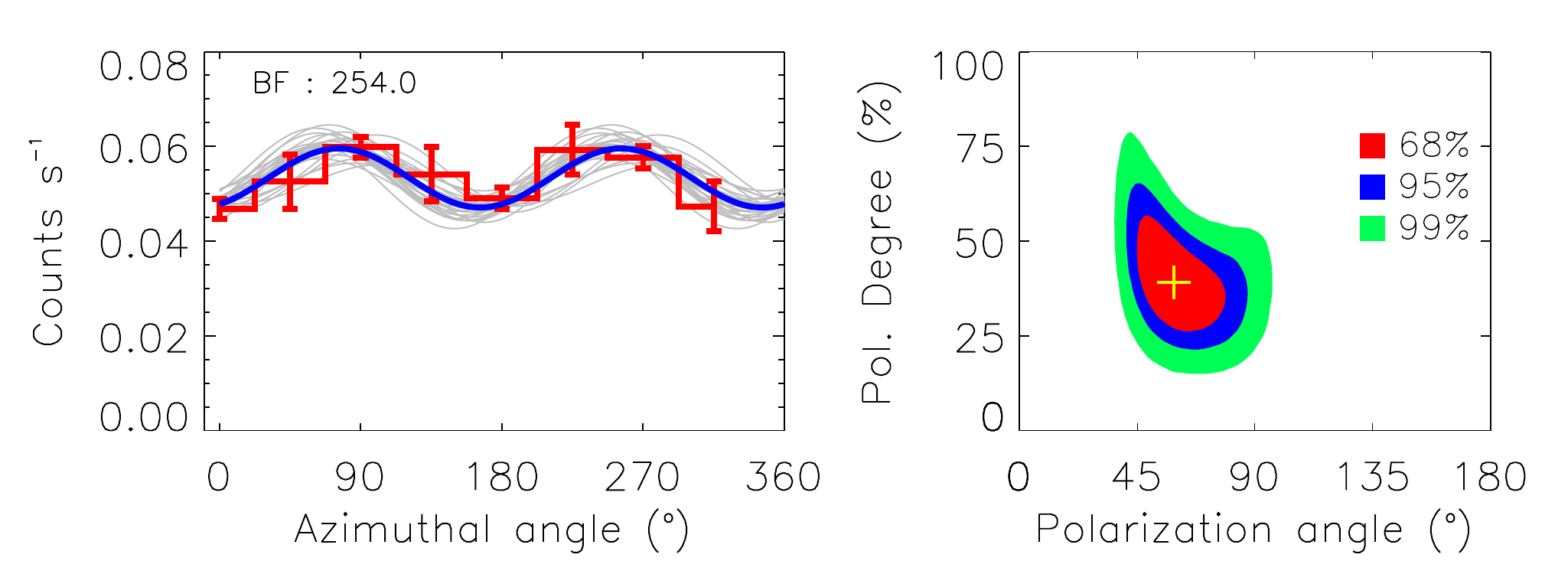}
     \caption{(f) HIMS2992: 230--380 keV}
    \end{subfigure}
	\caption{Results of polarization analysis with the left column showing  results for the whole observations in 100--380 keV and right column showing the results for the energy resolved polarization analysis for HIMS2992. Azimuthal Scattering Angle Distributions (ASAD) are shown on the left in each figure set. The sinusoidal fit is shown as a solid blue line, and 100 random MCMC iterations are shown as faint grey lines. The contour plots for polarization angle (in the sky plane with respect to the local north in anti-clockwise direction)  and polarization degree for 68, 95, and 99 \% confidence levels are shown in the right panels. X-ray emissions during PH5146 and SIMS4646 are unpolarized or polarized at low levels in 100--380 keV, whereas in HIMS2992, the emission is polarized at 23 \% with polarization angle 56$^\circ$. For HIMS2992, the first energy bin is unpolarized. The measured polarization degrees (and angles) for the other two energy bins are 26$\pm$6 \% (48$^\circ$) and 39$\pm$9 \% (59$^\circ$), respectively. For values of modulation factors in each case, please refer to Table \ref{tab:obs_res}.}
	\label{fig:results_pol}
\end{figure*}

We carried out a detailed polarization analysis for these three observations, following the steps described in \citet{vadawale17}. Details of the CZTI polarization measurement methodology can be found in supplementary section \ref{app:CZTI_pol}. The Azimuthal Scattering Angle Distribution (ASAD) and the contour plots for all the three observations are shown in Figure \ref{fig:results_pol} (a: PH5146, b: SIMS4646, c: HIMS2992).
The fitted modulations in ASAD for the PH and SIMS states are low and not constrained even though the estimated minimum detectable polarizations (MDPs) for PH5146 and SIMS4646 are low ($<$10 \%). We also estimate the Bayes factor, which provides a statistical confirmation of the detection of polarization by comparing a sinusoidal polarized model to an unpolarized constant model fitted to the data. The low Bayes factors ($<$3, as described in appendix \ref{app:CZTI_pol}), measured in both the cases, indicate no statistically significant detection of polarization in these two observations.    

Analysis of HIMS2992, on the other hand, shows statistically significant polarization, with measured polarization degree of 23$\pm$4 \% in 100--380 keV, implying greater than 5$\sigma$ detection for 1 parameter of interest at 68 \% confidence level. The observed polarization angle projected in the sky plane is 56$\pm$11$^\circ$, which agrees with the {\em INTEGRAL} results. The angle is $\sim$90$^\circ$ away from the $IXPE$ measured polarization angle in 2--10 keV. The contour plot on the right side of the figure shows that the polarization degree and the angle are well constrained at 68, 95, and 99 \%  confidence levels. The Bayes factor is also high ($\sim$733), confirming very high statistical significance.   

With such high detection significance for HIMS2992, we explored the energy dependence of the polarization. Figure \ref{fig:results_pol} (d), (e), and (f) show the modulation curves for HIMS2992 in three energy ranges: 100--175, 175--230, and 230--380 keV.
The signal in 100--175 keV is found to be unmodulated (Bayes factor $<$1). 
The signals at higher energies (175--230 keV and 230--380 keV), on the other hand, are found to be polarized (26$\pm$6 \% at 48$\pm$12$^\circ$ and 39$\pm$9 \% at 59$\pm$11$^\circ$ respectively) at $>$4$\sigma$ level (Bayes factor of 33 and 254, respectively). We measure upper limits of polarization for the data in the first energy bin (100--175 keV) of HIMS2992 and for the other two observations in 100--380 keV (see the rightmost column of table \ref{tab:obs_res}). In table \ref{tab:obs_res}, we also note the observed source, background counts, net exposures for the three observations and their expected minimum detectable polarizations (MDP) for 99 \% confidence level.  
 
\begin{figure}[h!]
	\centering
\includegraphics[scale=.45]{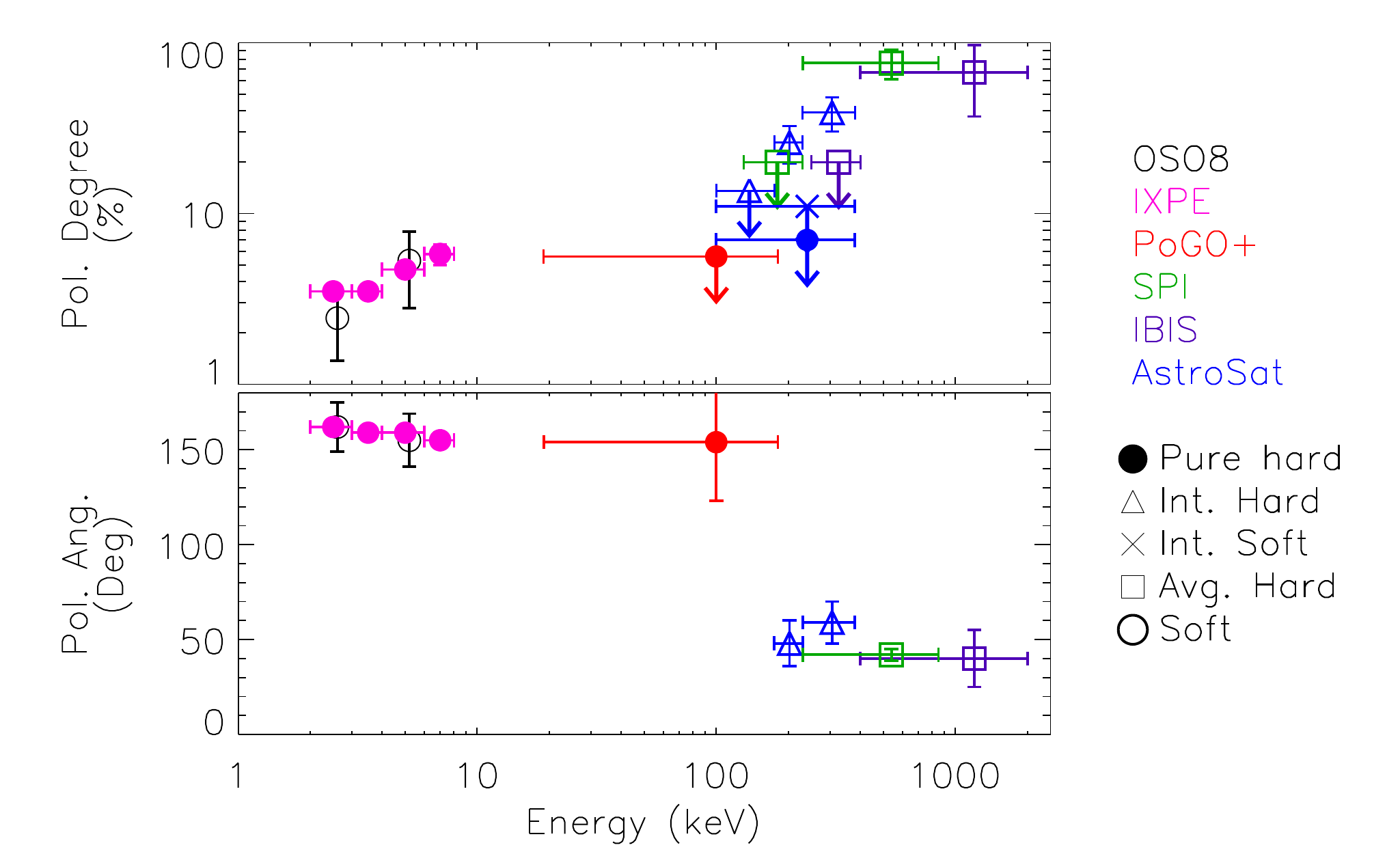}
\caption{Polarization degree (top panel) and angle (bottom panel) of Cygnus~X-1 in different spectral states (Pure hard, Intermediate hard, Intermediate soft, Averaged hard, and Soft states) from all the available measurements $-$ IXPE \citep{krawczynski22_ixpe}, IBIS \citep{laurent11} and SPI \citep{jourdain12} on board {\em INTEGRAL}, {\em OSO-8} \citep{long80}, PoGO+ \citep{chauvin18a} and {\em AstroSat}-CZTI. The polarization angle is measured from the local north to northeast in an anti-clockwise direction for all the instruments. When plotted against the observed energy of measurement, we see an apparent increase in the polarization degree and a swing in polarization angle at higher energies. The two distinct polarization angle distributions suggest different origins for the radiation below and above $\sim$200 keV.}
	\label{fig:results_all_exp}
\end{figure}

Figure \ref{fig:results_all_exp} shows the polarization degree and angle of Cygnus~X-1 in different spectral states (denoted by different symbols) from all available measurements till date (in different colors), including the {\em AstroSat} CZTI measurements presented here as blue data points.
Since the {\em INTEGRAL} data of Cygnus~X-1 encompass a few years of observation in total, we denote the results as averaged hard state, supposedly consisting of both pure hard and intermediate states.  
It can be seen that the CZTI measurements smoothly bridge the gap between the corona-dominated low energy ($<$ 100 keV) measurements of low polarization and the high polarization measured by {\em INTEGRAL}. 


\section{Spectroscopic analysis results} \label{sec:spec_results}
To investigate the spectral signatures of the polarization signal present in HIMS, we undertake spectroscopic analysis of the source for these three observations. 
The broadband X-ray spectrum of the source can be typically characterized as a combination of a thermal accretion disk and emission from a Comptonizing medium, with the Comptonized emission often showing structure; e.g two components with different optical depths \citep{Makishima2008PASJ...60..585M} and different electron temperatures \citep{Basak2017MNRAS.472.4220B} or hybrid distribution of electrons \citep{zdz2017MNRAS.471.3657Z} in the hard state of the source. 
Since the effect of the polarized component is limited to X-rays beyond 100~keV, we need to investigate the hard X-ray spectrum in detail with minimal dependence of model on the biases from lower energies. Therefore, for spectral analysis, we focus only on modeling the CZTI spectrum in 30--190~keV and do not include data in softer X-rays from the other two {\emph AstroSat} instruments: the Soft X-ray Telescope (SXT) and the Large Area X-ray Proportional Counter or LAXPC). 
\citet{torii2011PASJ...63S.771T} have investigated the hard X-ray spectrum (10--400~keV) of Cygnus~X-1 in the low hard state at multiple epochs and are able to model the emission with Comptonization with reflection from cold matter with {\tt compPS} \citep{compPS1996ApJ...470..249P}. Thus we use a similar formalism to model the CZTI spectrum in 30--190~keV. 
For each observation ID, we filter the raw event files following the standard CZTI data analysis pipeline procedure and generate clean event files. From the clean event files, we generate background subtracted source spectrum using an improved mask-weighting technique with updated calibration (Mithun et al in prep), which is implemented in \verb|cztbindata| 
module of CZTI data analysis pipeline version 3.0 and the associated 
CALDB\footnote{\url{http://astrosat-ssc.iucaa.in/cztiData}}. 
\begin{figure*}
	\centering
\includegraphics[width=.85\columnwidth]{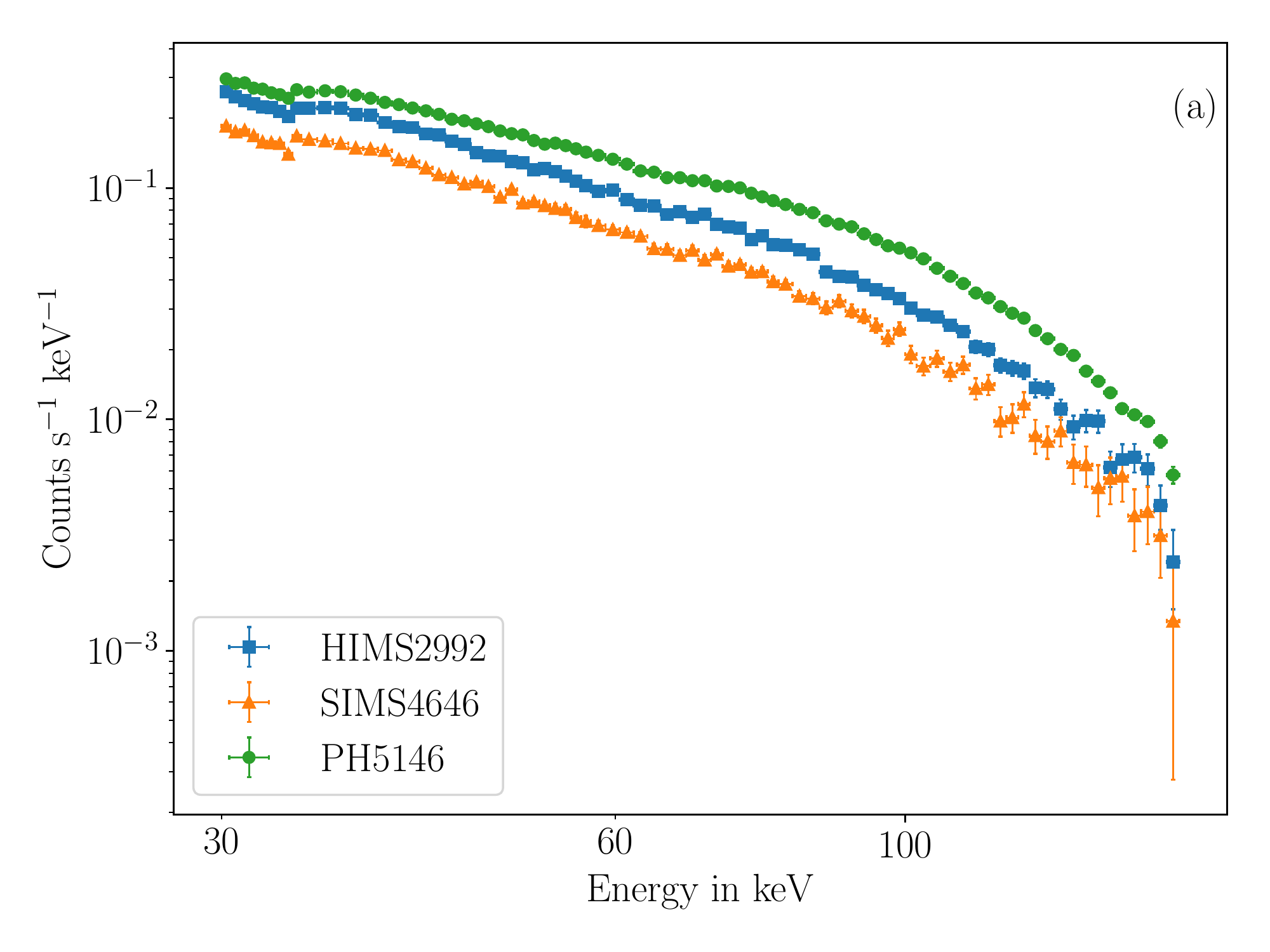}\includegraphics[scale=.53]{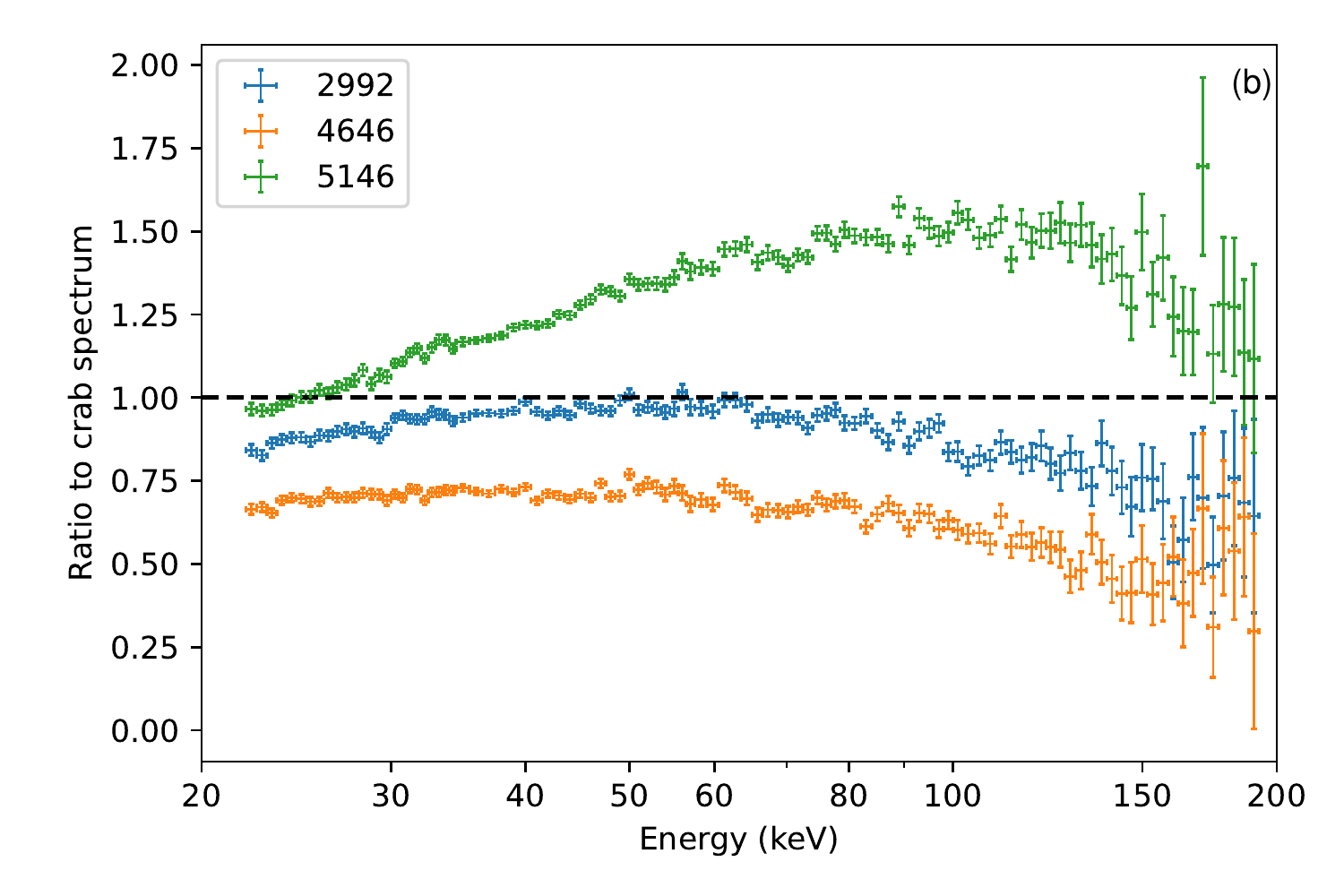} \\

\includegraphics[width=\textwidth]{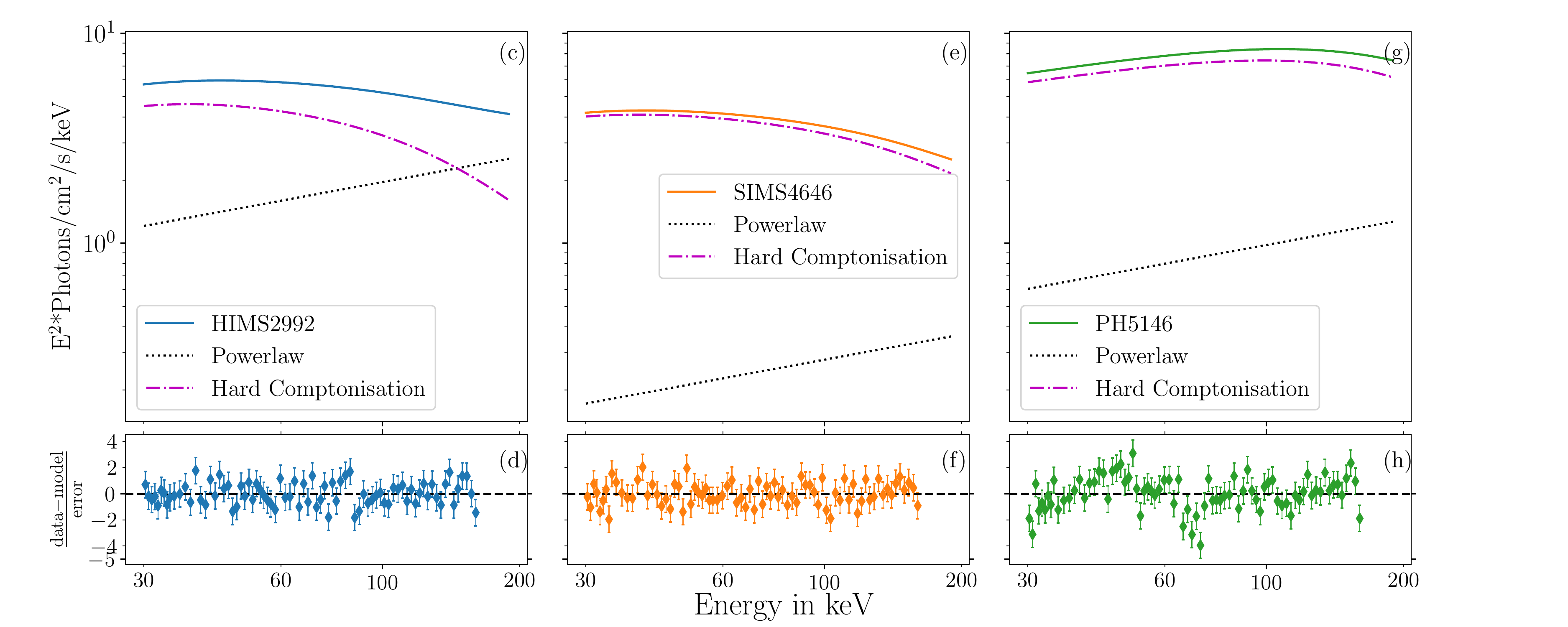}
\caption{Panel (a): Count spectra from the coded mask imaging for different epochs. Panel (b): The mask-weighted {\em AstroSat}-CZTI spectra normalized to Crab to highlight the spectral evolution across these epochs. 
Panels (c), (e) and (g) show the unfolded spectra fitted with \texttt{tbabs*(compps+powerlaw)}, for the three observations. Panels (d), (f) and (h) show the residuals from the fitting of spectrum with the model \texttt{tbabs*(compps+powerlaw)}.   The \texttt{compps} corresponds to Comptonization and \texttt{powerlaw} corresponds to an additional spectral component. Here the index ($\Gamma$) of the \texttt{powerlaw} component  is kept frozen in the analysis (see text for more details). The complete set of fitted spectral parameters for both the models: \texttt{tbabs*(compps+powerlaw)} and \texttt{tbabs*compps} is given in Table \ref{tab:spec_pars} in appendix \ref{app:spec_tab}.
}
	\label{fig:spectra}
\end{figure*}

In Figure~\ref{fig:spectra} panel (a), we show the count spectra of the source in the three states (blue: HIMS2992, orange: SIMS4646, green: PH5146) whereas the inherent differences in the shape of the spectrum in the three states are shown in panel (b) by computing the ratio of the respective spectrum to Crab spectrum. This removes the instrumental effects and allows for comparison of the spectral slopes in a model-independent manner. The PH5146 state has the highest flux and shows the characteristics of thermal Comptonization spectra with a  cutoff at $\sim$100 keV. In the other two states (HIMS2992 and SIMS4646), there is an indication of the cutoff energy being lower and the emergence of a power law component. Since we are restricted to energies above 30 keV, for spectral modeling, we fix all parameters that affect below this energy (i.e. disk temperature and reflection parameters) to their typical values \citep{Makishima2008PASJ...60..585M} and assume a spherical geometry of the Comptonizing medium \citep{Makishima2008PASJ...60..585M, torii2011PASJ...63S.771T}. Thus only parameters which we constrain are the optical depth of the medium, electron temperature, and the normalization. 
We find that all the three observations can be described adequately with a hard Comptonization ($\chi^2$ of 58.4, 58.52 and 133.4 for 76 degrees of freedom in HIMS2992, SIMS4646, and PH5146 respectively) with the parameters consistent with those reported by \citet{Makishima2008PASJ...60..585M, torii2011PASJ...63S.771T} with the electron temperature in case of PH roughly similar to typical reported electron temperature using hard X-ray observations \citep{Basak2017MNRAS.472.4220B}. The fitted parameter values are given in Table \ref{tab:spec_pars} in appendix \ref{app:spec_tab}. The electron temperature in HIMS is higher ($\sim$170~keV) than that observed in SIMS ($\sim$151~keV) or PH  ($\sim$87~keV) while the optical depths follow a reverse trend. Confidence intervals of the parameters are determined by Markov Chain Monte Carlo sampling of the parameter space using 50 walkers running for 5000 steps (after burning initial 2000 steps before convergence). To estimate the effect of Comptonizing medium geometry on the estimated parameters, we switched the `geometry' to cylindrical and slab geometry for HIMS2992 in different tests. 
Like before, we estimated the parameters of the Comptonizing medium allowing the relevant parameters to be free. In both cases, we estimate similar electron temperatures and optical depths with H/R$\approx$2 in the case of cylindrical geometry and a covering fraction of $\approx$1 in slab geometry. The change in geometry doesn't result in a better fit as the change in the $\chi^2$ is negligible for a reduction in the degrees of freedom by 1. Due to a simpler model description of the spherical geometry, we conduct further spectral modelling assuming a spherical geometry for the Comptonizing medium.

\citet{laurent11} have reported the presence of a powerlaw-like component in the \textit{INTEGRAL} spectrum of Cygnus~X-1 in its hard state. We test for the presence of a similar component in the CZTI spectra by including a {\tt powerlaw} with a fixed slope and a variable normalization. The addition of the component does not change the fit statistics (reduced $\chi^2$ is $\lesssim$1) but in the case of HIMS observation, it causes a significant change in the electron temperature. We keep the slope of the \texttt{powerlaw} component tied at 1.6 as it is representative of the typical synchrotron jet observed in the INTEGRAL observation \citep{laurent11} and since we only want to test the presence of the component, computing the significance of the component by estimating its normalization is sufficient.  The model decomposition for the individual observations 
(for the second model: \texttt{tbabs*(compps+powerlaw)}) is shown in Figure~\ref{fig:spectra} (c,d: HIMS2992; e,f: SIMS4646; g,h: PH5146). The model parameters are noted in Table~\ref{tab:spec_pars} in appendix~\ref{app:spec_tab}.
We note that the spectral modeling of the HIMS state allows inclusion of a \texttt{powerlaw} component with reasonable constraints on its normalization, though with the inclusion of the \texttt{powerlaw} component, the electron temperature reduces by a factor of 2.7 and is roughly similar to the electron temperature observed in PH. The parameters after the inclusion of the \texttt{powerlaw} component are roughly similar to the ones estimated using other formalisms (e.g using \texttt{eqpair}  \citep{coppi1999ASPC..161..375C} in \cite{zdz2012MNRAS.423..663Z}, \texttt{nthcomp} \citep{zdz1996MNRAS.283..193Z, zycki1999MNRAS.309..561Z} in \cite{Basak2017MNRAS.472.4220B}). The differences may arise due to the assumptions but given the similarity of the parameters, it would indicate a similar picture being portrayed. The SIMS and PH allow inclusion of the power-law component, but with much lower normalization and is consistent with zero within a few standard deviations. 
Based on the spectroscopic analysis, we conclude that there is a degeneracy in the spectral information in HIMS2992 with both Comptonization and Comptonization + Powerlaw models being able to describe the spectrum. The latter configuration aligns better with the polarization results if we assume that the \texttt{powerlaw} component is the main contributor to the observed polarization in $>175$~keV. The measured relative contributions of the \texttt{powerlaw} (1.9$\times$10$^{-9}$ erg/cm$^2$ in 100--175 keV, 1.1$\times$10$^{-9}$ erg/cm$^2$ in 175--230 keV, and 2.4$\times$10$^{-9}$ erg/cm$^2$ in 230--380 keV) to the total flux (4.2$\times$10$^{-9}$ erg/cm$^2$ in 100--175 keV, 1.8$\times$10$^{-9}$ erg/cm$^2$ in 175--230 keV, and 2.9$\times$10$^{-9}$ erg/cm$^2$ in 230--380 keV) are consistent with the flux contributions expected from the observed polarization results within 1$\sigma$ scatter of each other, assuming a 40 \% maximum polarization from the synchrotron flux (more details in appendix \ref{app:spec_tab}). In appendix \ref{app:spec_tab}, we also extract the spectrum of the Compton events and model the Compton
spectra simultaneously with the CZTI mask weighted spectra to show its consistency with the mask-weghted spectra for all the three
observations. We find these results (reduced $\chi^2$ and \texttt{powerlaw} fluxes) to be similar when we use a slightly steeper \texttt{powerlaw} index (1.8) for HIMS2992. This further corroborates evidence of additional spectral component above 100 keV mainly during the high polarization substate. We would like to mention here that the present result suggests a  hard (Gamma 1.6 or 1.8) component above 100 keV, not generally seen in the low hard state of the source \citep{zdz2012MNRAS.423..663Z}. This could be because of the different spectral sub-states where high polarization is seen.  Our current data do not enable examining other more complicated models, e.g., hybrid Comptonization models, which might also be able to describe the additional spectral component at higher energies \citep{Poutanen2009,malzac2009}. 
High polarization degree around 50\% in hybrid Comptonization, however, requires stringent conditions on the Comptonization geometry and high relative population of the non-thermal electrons. 
Wide band sensitive spectral data and polarization in future will be useful to reexamine these different models in different substates and develop deeper theoretical understanding.  


\section{Summary and Discussions} \label{sec:discussion}
In this paper, we report new polarization measurements of Cygnus~X-1 using the {\em Astrosat}-CZTI instrument in 100--380 keV. Polarization measurements were done in three different spectral states $-$ pure hard (PH), intermediate hard (HIMS), and intermediate soft states (SIMS). In the PH and SIMS states, we did not see any evidence of polarization (upper limit of $\sim$10 \%). However, the HIMS state was seen to have polarized emission with a fraction of 22 \%, measured with more than 5$\sigma$ significance at an angle 
around 236$^\circ$ (local north to east in anti-clockwise direction). Energy resolved analysis shows that the polarization increases with energy from no polarization at energies $<$175 keV to $\sim$40 \% polarization at higher energies. The high polarization reported here is in an energy range (175-380 keV) where CZTI does not have imaging information. Though we are confident that the measurements below 190 keV are from Cygnus X-1 (based on coded mask imaging)  and that there are no contaminating sources near Cygnus X-1 based on the Swift-BAT data (below 195 keV), we cannot completely rule out the remote possibility that a source unrelated to Cygnus X-1 but having the peculiar property of having low flux below 200 keV and high flux and polarization above this energy, is contaminating our results.
 
It has been known for a long time that the spectral shape observed at high energies for Cygnus~X-1 requires multiple components. The two distinct polarization angles measured at low and high energies (see Figure \ref{fig:results_all_exp}) strongly indicate the existence of a distinct spectral component at high energies, with an origin different from that in the putative corona 
\citep{bel06_cgx1,Jourdain14_cgx1,rahoui11_cgx1}. 
The high polarization seen here strongly links this component to synchrotron radiation in an ordered magnetic field, possibly from the base of the jet. Further, finding high polarization confined only to the HIMS state provides further clues to the origin of jets in Cygnus~X-1.

In the hard state, the different sub-classes (PH, HIMS, and SIMS) are configurations of the accretion disk dictated by accretion rate, location of disk truncation, and the strength of outflows and jets. 
HIMS state is fascinating because, in this spectral state, maximum radio flux variation has been observed \citep{Lubinski20}, and strong jets are expected to be formed. We also detect high polarization in this state. The evolution of polarization degree with energy in this state along with the spectral analysis results, therefore, favors a scenario where the coronal and jet emission mechanisms co-exist in 100--380 keV energy range and intersect around 200 keV. 
In the PH and SIMS spectral states, on the other hand, there is no evidence of polarization from the jet component, either in the energy-integrated or in the energy-resolved analysis. However, it is to be noted that steady radio emission is seen in both these states, although the flux and its variation are found to be low \citep{Lubinski20}. 
This suggests that the jet component may be present in all three states with similar polarization properties, but in the PH and the SIMS states, the X-ray emission is dominated by the Corona all the way to $\sim$400 keV. 

Synchrotron process in a highly ordered magnetic field in a jet represents a probable way to produce highly polarized emission. However, as emphasized by \citet{zdziarski14},  high polarization levels in X-rays (e.g. seen in {\em INTEGRAL}) require extreme conditions that are unlikely to occur in a steady-state jet.
Recently, \citet{russell14_cgx1} attempted multi-wavelength SED modeling of Cygnus~X-1 spectral and polarimetric data (radio, IR, optical, and  {\em INTEGRAL} data in X-rays), based on synchrotron emission in an ordered magnetic field of a steady-state jet (with a simplified the magnetic field topology composed of straight lines). They failed to explain the high-energy polarization angles ($\sim$60$^\circ$ away from the jet) reported by {\em INTEGRAL}, while the optical and IR emissions are polarized in the direction of the jet. One expects the intrinsic polarization angle to be wavelength independent when a single electron population in an optically thin jet is responsible for emission across the electromagnetic spectrum.
These contradictions suggest an alternative possibility that the observed high polarization may result from some peculiar transient phenomena occurring mainly in the HIMS state of the source. 

In transient black hole binaries, it is observed that sources traverse well-defined paths of state transitions starting from the low hard state (with the indication of a steady jet as evidenced by the radio emission) and then make a state transition to the soft State (Belloni et al. 2005). This transition, quite often, passes through the HIMS state, and it is even suggested that during this transition, the source passes through a specific ‘jet-line’ where super-luminal jet ejections are observed to take place (Fender et al. 2004). Cygnus X-1 is a high-mass X-ray binary, and, in contrast to the black hole transients, it makes slow transitions and spends several days in each sub-state (though superluminal jet emission episodes are not seen in Cygnus X-1). It is quite conceivable that we are seeing the ‘jet-line’ transition in slow motion during the HIMS state of Cygnus X- 1. Hence, many of the assumptions of the steady-state jets, which failed to explain the high polarization and the different PA, may not be valid during a transient jet. The constraints presented in this work for any such transient jet phenomena are (assuming synchrotron emission from a plasma in an ordered field): a) very high polarization, b) magnetic axis almost along the jet axis (note that the measured polarization angle is close to 90 degrees to the jet axis). Observationally, a detailed time-resolved multi-wavelength observation of Cygnus X-1 in the HIMS state would be instrumental in understanding this enigmatic source.

\begin{acknowledgments}
This publication uses data from the {\em AstroSat} mission of the Indian Space Research 
Organisation (ISRO), archived at the Indian Space 
Science Data Centre (ISSDC). CZT-Imager is built 
by a consortium of institutes across India, 
including the Tata Institute of Fundamental 
Research (TIFR), Mumbai, the Vikram Sarabhai Space 
Centre, Thiruvananthapuram, ISRO Satellite Centre 
(ISAC), Bengaluru, Inter University Centre for 
Astronomy and Astrophysics, Pune, Physical 
Research Laboratory, Ahmedabad, Space Application 
Centre, Ahmedabad. Contributions from the vast 
technical team from all these institutes are 
gratefully acknowledged. Specifically, we would 
like to thank M. K. Hingar, A. P. K. Kutty, M. H. 
Patil, S. Sinha and Y. K. Arora (TIFR) for the CZT-
Imager hardware fabrication; and K. S. Sarma, K. 
H. Navalgund, R. Pandiyan and K. Subbarao (ISAC) 
for project management and mission operation. The 
continued support from M. Annadurai and A. S. 
Kirankumar is gratefully acknowledged. 
\end{acknowledgments}

\appendix

\section{Determination of spectral states} \label{app:spec}
CZTI data consists of a time-tagged event list with a time resolution of 20 $\mu$s which include the information of the CZTI quadrant, CZT detector module ID, pixel number, and PHA value for each event. CZTI data reduction pipeline takes this event list
as an input and generate standard data products like light curves and spectra. For generation of background subtracted spectrum and light curves, CZTI analysis pipeline makes use of mask-weighting technique where the background is measured simultaneously considering the pixels' open fractions and effective areas. 
In order to identify the spectral state class of ID2992, ID4646, and ID5146, we followed the same technique prescribed by \citet{Lubinski20}. 
They have done a detailed spectral analysis of Cygnus~X-1 using {\em INTEGRAL} data spanning over fifteen years. and categorised the states into hard and soft regimes based on the hard X-ray flux in 22--100 keV: hard state for flux above 75$\times10^{-10}$ erg cm$^{-2}$ s$^{-1}$ and soft state for flux below this value. Each regime is further categorised into pure, transitional and intermediate, totaling six states -- pure hard (PH, $\Gamma \le 1.78$), transitional hard (TH, $1.78\le \Gamma \le 1.93$ ), hard intermediate (HIMS, $1.93 \le \Gamma \le 2.29$), soft intermediate (SIMS, $1.93 \le \Gamma \le 2.29$), transitional soft (TS, $2.29 \le \Gamma \le 2.65$), and pure soft (PS, $\Gamma > 2.65$) states based on the clustering of the data in spectral index and flux density diagram.

Spectral analysis similar to the \citep{Lubinski20} requires analysis of hourly or sub-hourly data (0.5--2 hour). Since each orbit of the {\em AstroSat} lasts for $\sim96$ minutes, we proceeded with spectral analysis of orbit-wise data.  
Each of the three observations is divided into separate orbit-wise cleaned event files by applying the orbit-wise Good Time Interval (GTI), which is obtained using a program written in Interactive data Language (IDL). We exclude the South Atlantic Anomaly (SAA) regions for each orbit. 
The orbit-wise event files are then used to obtain the spectral and response files from the standard CZTI data reduction pipeline. 
The spectral analysis is carried out using the {\tt xspec} \citep{arnaud1996xspec} for each orbit files. The spectrum for all the observations is fitted with an {\tt powerlaw} model. The spectral indices for the orbits in 30--100 keV and the computed model flux in 22--100 keV obtained from the spectral fitting are plotted in Figure \ref{fig:states} for the three observations. The average values of the spectral indices and flux are given in table \ref{tab:obs_res}. Based on these values, we determine that ID5146 is a PH state, ID2992 is a HIMS state, and ID4646 is a SIMS state.  

\section{X-ray polarimetry with CZT-Imager} \label{app:CZTI_pol}
In the CZTI, polarization is estimated from the azimuthal scattering angle distribution (ASAD) of the Compton scattered photons \citep[see][for the Compton scattering technique details]{denis22,chattopadhyay21_review,lei97}. The CZTI consists of a large pixelated detector plane (geometric area of  976 cm$^2$) with pixel size of 2.5 mm $\times$ 2.5 mm and 5 mm thickness, and possesses considerable Compton scattering efficiency above 100 keV, making it suitable for Compton scattering polarimetry in 100--380 keV. The on-board electronics preserve simultaneous multi-pixel events and enable time-tagged transmission of individual events, thus providing polarization information on a routine basis. Here we brief the polarization analysis steps \citep{vadawale17}. 

\subsubsection*{Selection of Compton events}
The first step of the polarization analysis is to select valid Compton events. For each of the three individual observations, we removed the intervals of high background before and after the South Atlantic Anomaly passage in each orbit. We also removed all events from pixels that were classified as noisy or spectroscopically bad. Then, we extracted the double pixel events satisfying the Compton criteria, e.g., detected within 20 $\mu$s time window in two adjacent pixels and the energetics of the events satisfy Compton kinematics. For details, see \citet{chattopadhyay14,vadawale15}. The Compton events are then used to obtain the source ASAD.

\subsubsection*{Background subtraction}
It is important to consider an appropriate blank sky observation for the measurement of background. Since the CZTI mask and other support structures become increasingly transparent at energies beyond 100 keV, a bright X-ray source at a large off-axis angle, even up to 80$^\circ$, can interfere with the true background. The blank sky observations were taken from a region where the Crab and Cygnus~X-1 are out of the open field of view of the CZTI. We also need to consider the effect of the earth X-ray albedo, which constitutes a large fraction of the hard X-ray background in the low earth orbit. Since CZTI is at the corner of the spacecraft with most instruments present only at one side of CZTI and albedo background comes from one side of the spacecraft, this may lead to an asymmetry in the background azimuthal scattering angle distribution. In order to minimize this effect, the blank sky observations were selected such that the relative orientation of the spacecraft during the background measurement is the same ($\pm$5$^\circ$) as that during the source measurement. The same data cleaning process and Compton criteria are implemented to generate the background ASAD. 

Prior to the background ASAD subtraction, an important point to consider is that the background count rate changes within the duration of observation with a stable periodic nature. This results from the inclined orbit of {\em AstroSat} where some of the orbits pass through the outskirts of the South Atlantic Anomaly giving an increase in count rate when the spacecraft is in these regions of the orbit. Because of the rotation of the earth, this phase of high count rate reappears every $\sim$24 hours and this has been seen in other {\em AstroSat} instruments also \citep{antia22} apart from CZTI \citep{kumar21,kumar22}. To correct for this effect, we try to match the phases of orbital variation of the count rate during background and Cygnus~X-1 observations using a cross-correlation method \citep{kumar21} and identify the common or phase-matched regions. These phase matched regions are used to correct for the long-term variation in data before background subtraction.

\subsubsection*{Modulation curve fitting}
In the next step, we fit the ASAD to obtain polarization degree and angle. Because of the non-uniformity in the solid angles subtended by the surrounding edge and corner pixels to the central scattering pixel, we see an unequal count rates in the edge and corner pixels, which is first corrected by normalizing it with an ASAD for 100 \% unpolarized radiation in the same energy range. For Gamma-ray bursts, this is typically obtained from the Geant4 \citep{agostinelli03} Monte Carlo simulation \citep{chattopadhyay19,chattopadhyay22_grb}. However, for ON-axis sources, the unpolarized ASAD is best obtained from the observed source azimuthal distribution by averaging the edge and corner pixels separately \citep{vadawale15,vadawale17}. The geometry-corrected modulation curves are fitted by a sinusoidal function, $A\cos{2(\phi - \phi_0 + \pi/2)}+B$, to estimate the polarization angle in the detector plane ($\phi_0$) and the modulation amplitude ($\mu=A/B$).

Errors on the raw ASAD of source and background observations are computed individually based on counting statistics and propagated to calculate the errors on the background-subtracted ASAD.
To estimate the values of the fitting parameters (A, B, $\phi_0$) and the uncertainties on them, we perform MCMC 
simulations for a large number (1 million) of iterations. For each iteration, the posterior probability is estimated based on randomly sampled model parameter values. At the end of the evolution chain, while the modulation factor and polarization angle are estimated from the best fitted 
values of the parameters ($A, B$ and $\phi_0$), uncertainties 
on them are computed from the distribution of the posterior probabilities
of the parameters. The polarization degree for each of these observations are estimated by normalizing the fitted $\mu$ by the modulation factor expected for 100 \% polarized radiation ($\mu_{100}$), which is obtained from Geant4 simulations where an identical process for the selection of Compton events is followed. To confirm that observation is statistically polarized, we estimate the Bayes factor for the sinusoidal model (M$_1$, for polarized photons) and a constant model (M$_2$, unpolarized photons) as the ratio of marginal likelihoods of M$_1$ to M$_2$ \citep[for more details, see][]{chattopadhyay19}.  
In the cases where the Bayes factor is greater than 3, we estimate polarization degree and angle from the fitted parameters (for example, for HIMS2992, the Bayes factor is $>$3 in the full energy range and in 175--230 keV and 230--380 keV). If the Bayes factor is estimated to be less than 3 (e.g., PH5146, SIMS4646, and HIMS2992 in 100--175 keV), we estimate polarization upper limit.

\section{Spectral fit results} \label{app:spec_tab}
The spectral data for the three observations (HIMS2992, SIMS4646, and PH5146) were fitted with two different models - \texttt{tbabs*compps} and \texttt{tbabs*(compps+powerlaw)}. In Table \ref{tab:spec_pars}, the fitted parameter values are summarized for the models.

\begin{table}[h!]
    \centering
    \caption{Key spectral parameters for the models \texttt{tbabs*compps} and \texttt{tbabs*(compps+powerlaw)}, where \texttt{compps} corresponds to Comptonization and \texttt{powerlaw} corresponds to an additional spectral component.} \label{tab:spec_pars}
    \begin{tabular}{lll|rrr}
\hline
         Component & Parameter & Unit & 2992 (HIMS) & 4646 (SIMS) & 5146 (PH) \\ \hline
         \multicolumn{6}{c}{\texttt{tbabs*compps}} \\ \hline
         \texttt{tbabs} & N$_{\rm{H}}$ & (10$^{22}$ cm$^{-2}$)      & \multicolumn{3}{c}{0.66$^\dagger$} \\ \hline
         \texttt{compps} & kT$_e$ & keV & $176\pm20$ & $151_{-6}^{+8}$ & $87_{-2}^{+2}$ \\
                        & $\tau$ & & $0.50_{-0.08}^{+0.10}$ & $0.58_{-0.05}^{+0.04}$ & $1.40\pm0.04$ \\
                        & Norm & 10$^5$ & $38\pm5$ & $1.04\pm0.07$ & $1.31_{-0.03}^{+0.04}$ \\ \hline
         & $\chi^2$/dof & & 58.4/76 & 58.52/76 & 133.4/76 \\ \hline
         \multicolumn{6}{c}{\texttt{tbabs*(compps+powerlaw)}} \\ \hline
         \texttt{tbabs} & N$_{\rm{H}}$ & (10$^{22}$ cm$^{-2}$)      & \multicolumn{3}{c}{0.66$^\dagger$} \\ \hline
         \texttt{compps} & kT$_e$ & keV & $65\pm3$ & $103_{-5}^{+7}$ & $81_{-4}^{+3}$ \\
                        & $\tau$ & & $1.5_{-0.09}^{+0.06}$ & $0.78_{-0.06}^{+0.07}$ & $1.48_{-0.12}^{0.01}$ \\
                        & Norm & 10$^5$ & $22\pm2$ & $0.55\pm0.05$ & $1.21_{-0.01}^{+0.10}$ \\ \hline
        \texttt{powerlaw} & $\Gamma$ & &\multicolumn{3}{c}{1.6$^\dagger$} \\
                        & Norm & & $0.31\pm0.04$ & $0.07_{-0.03}^{+0.04}$ & $0.15_{-0.12}^{+0.09}$ \\ \hline
        & $\chi^2$/dof & & 56.9/75 & 58.47/75 & 132.5/75 \\ \hline
    \end{tabular}
    \begin{flushleft}
    Note: The errors reported in the table are 1$\sigma$ confidence interval. \\
    $^\dagger$ The parameter was kept frozen
    \end{flushleft}
\end{table}

\begin{figure*}[hb!]
	\centering
\includegraphics[scale=.49]{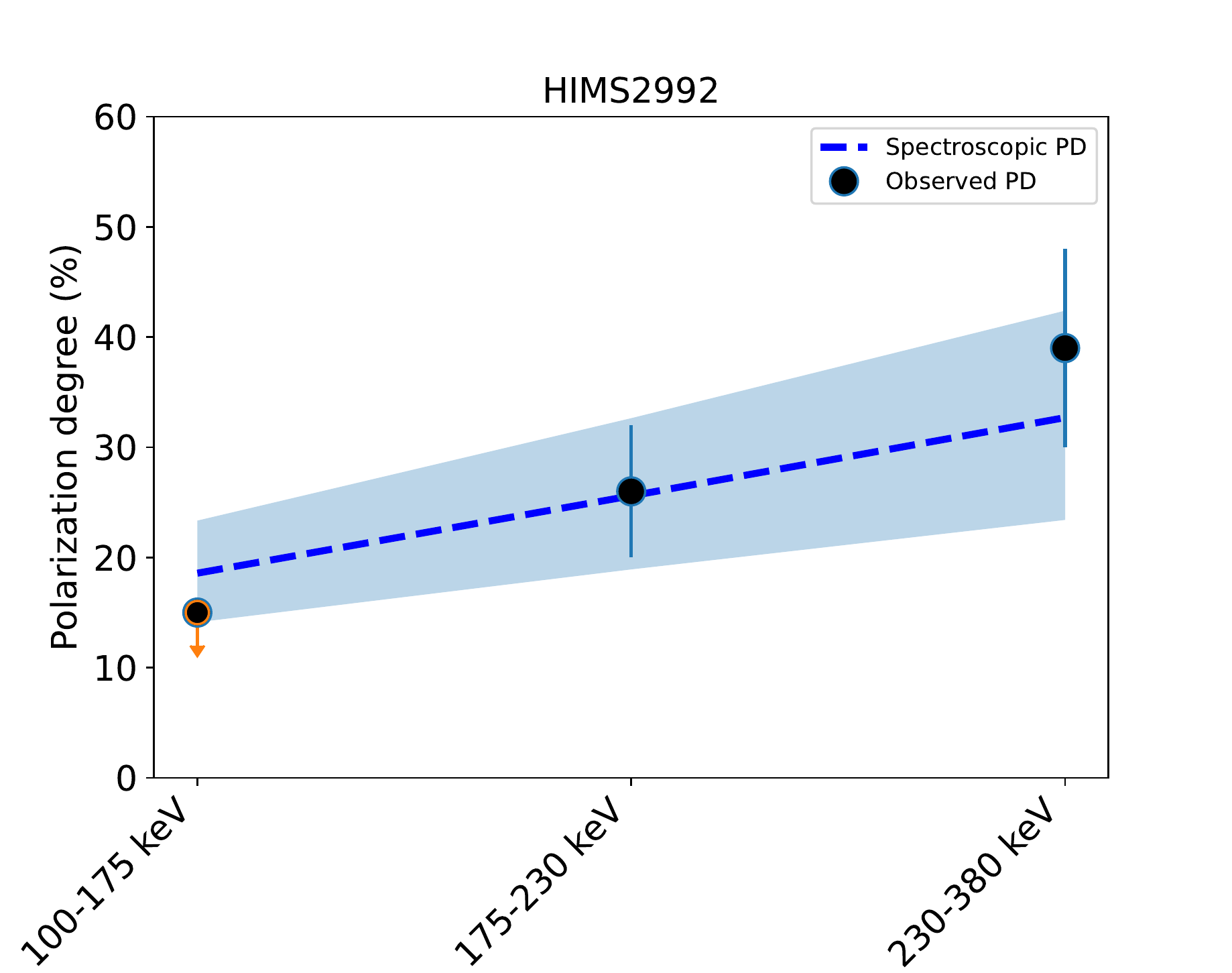}
\caption{Comparison between the observed polarization degree for HIMS2992 and the expected polarization degree computed from the measured power law (index of -1.6) flux contributions as a function of energy. Here we assumed a maximum 40 \% of synchrotron polarization degree. The error bars are computed at 1$\sigma$ level.}
	\label{fig:pf_comparison}
\end{figure*}
From the fitted power law and Comptonization parameters for HIMS2992, we computed the contribution of the power law component to the total flux in 100--175, 175--230, and 230--380 keV, respectively. Assuming a synchrotron origin of the power law, we computed the expected polarization degree as a function of energy assuming a maximum of 40 \% synchrotron polarization and compared it with the observed polarization degree as shown in Figure \ref{fig:pf_comparison}. The measured and predicted polarization degree agree well within 1$\sigma$ deviation of each other. 

\subsubsection*{Verifying the spectral properties with Compton spectroscopy}
To investigate if the spectral modeling of the source in different states of the source also extends to the energy range in which polarization is estimated (i.e. 175--380~keV), we extract the spectrum of the Compton events and model the Compton spectra simultaneously with the CZTI mask weighted spectra. 
Unlike CZTI coded mask spectra, there is no simultaneous background measurement available for the Compton events. The background is, therefore, estimated from the blank sky observations and then scaled which can lead to systematic uncertainties. This adds to the complexity of CZTI Compton spectroscopy, particularly for fainter and variable sources and sources with complicated spectral features like Cygnus~X-1. The details of the steps involved can be found in \citet{kumar22}. 

\begin{figure}
    \centering
    \includegraphics[width=\columnwidth]{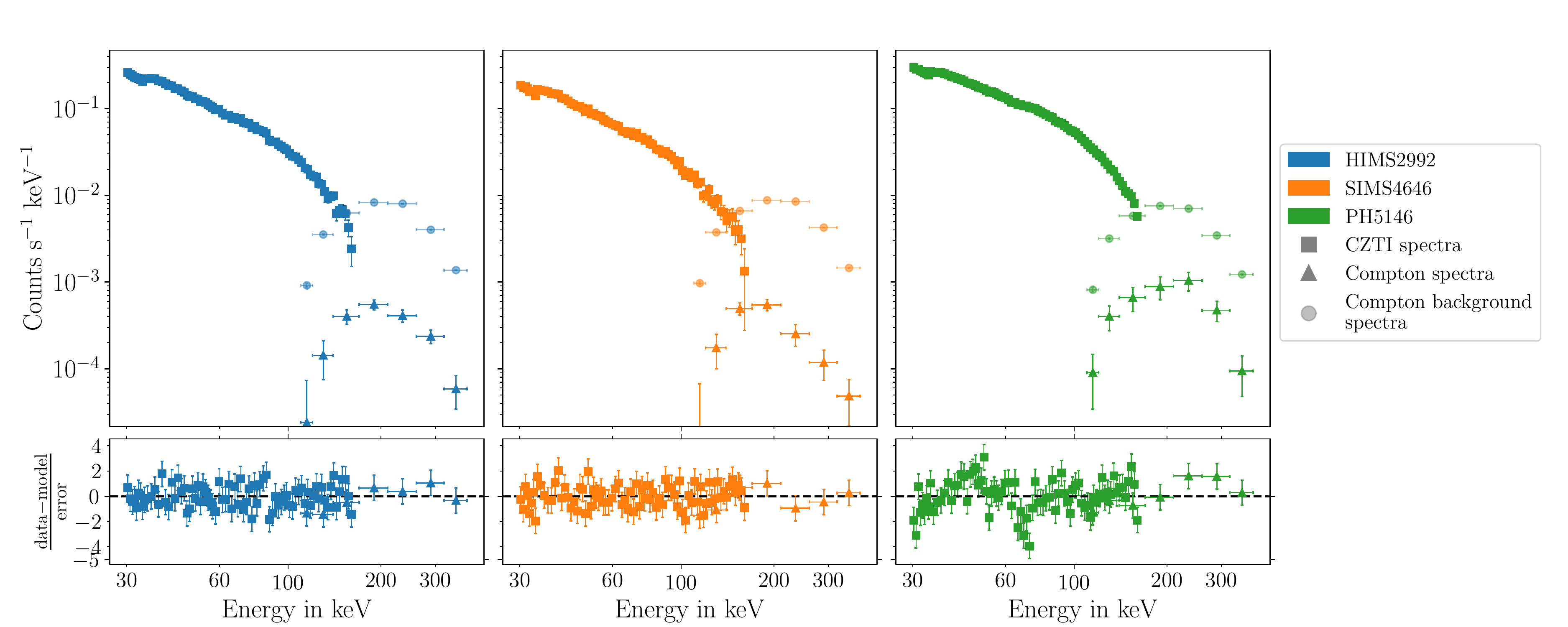}
    \caption{Count spectra (top panels) and the residual to the best fit model (bottom panels) from CZTI for various observations (different colors). The spectrum from mask weighting analysis is shown with squares (and is background subtracted) and the Compton double events spectrum is shown with triangles. The background spectrum for the Compton double event spectrum is estimated from a separate observation and is shown in circles.   }
    \label{fig:compt_spec}
\end{figure}

Since we are primarily interested to investigate the consistency of the Compton spectra with the respective CZTI coded mask spectrum, we keep the spectral parameters fixed to ones reported in table~\ref{tab:spec_pars}, and only allow for a free cross-normalization constant. We find that in all the cases, the  Compton spectrum is consistent with the model from the respective mask weighted CZTI spectrum with the cross-normalization constant $\approx1$. 
In case of observations ID2992 and ID4646, we find that the $\chi^2$ contributed towards the spectral fitting is low ($\sim$7 for 7 bins). However, in the case of ID5146, an additional 3\% systematic error was included to mitigate slight uncertainty in the background spectra. Because of these uncertainties, the Compton spectra can not be used to constrain the spectral parameters more accurately, but this analysis clearly demonstrates that Compton spectra are consistent with the mask-weghted spectra of CZTI for all the three observations. 
We depict the Compton double event spectra and CZTI mask-weighted spectra and their residuals to the best-fit model in Figure~\ref{fig:compt_spec}. 

\end{document}